**Scalable Substrate Development for Aqueous Biological Samples for Atom Probe Tomography**

Eric V. Woods[1, *], Se-Ho Kim[1,2], Ayman A. El-Zoka[1,3], L.T. Stephenson[1,4], and B. Gault[1,5, *]

1. Max-Planck-Institut für Eisenforschung, Düsseldorf, Germany

2. *now at* Department of Materials Science and Engineering, Korea University, Seoul 02841, Republic of Korea

3. *now at* Department of Materials, Royal School of Mines, Imperial College London, London, UK.

4. *now at* Australian Centre for Microscopy and Microanalysis, The University of Sydney, Sydney, NSW, Australia

5. Department of Materials, Royal School of Mines, Imperial College London, London, UK.

*Corresponding Authors: e.woods@mpie.de, b.gault@mpie.de

# Abstract

Reliable and consistent preparation of atom probe tomography (APT) specimens from aqueous and hydrated biological specimens remains a significant challenge. One particularly difficult process step is the use of a focused ion beam (FIB) instrument for preparing the required needle-shaped specimen, typically involving a "lift-out" procedure of a small sample of material. Here, two alternative substrate designs are introduced that enable using FIB only for sharpening, along with example APT datasets. The first design is a laser-cut FIB-style half-grid close to those used for transmission-electron microscopy, that can be used in a grid holder compatible with APT pucks. The second design is a larger, standalone self-supporting substrate called a "crown," with several specimen positions that self-aligns in APT pucks, prepared by electrical discharge machining (EDM). Both designs are made nanoporous, to provide strength to the liquid-substrate interface, using chemical and vacuum dealloying. We select alpha brass a simple, widely available, lower-cost alternative to previously proposed substrates. We present the resulting designs, APT data, and provide suggestions to help drive wider community adoption.

# 1   Introduction

Analysis of materials in their native, hydrated form with cryogenic sample preparation has revolutionized different aspects of surface science and materials characterization. The advent of cryogenic transmission electron microscopy (cryo-TEM), which received the 2017 Nobel Chemistry Prize (Nannini, 2017, Dubochet *et al.*, 1988), advanced the imaging of proteins in their native state, as well as other biomolecules, macromolecular complexes, and viruses (Walther *et al.*, 1992, Mahamid *et al.*, 2016, Schaffer *et al.*, 2017, Tonggu & Wang, 2020). However, cryo-TEM has historically been limited by sample preparation techniques (Doerr, 2016, Cheng *et al.*, 2015). New developments in cryo-sample preparation methods (Klumpe *et al.*, 2022, Parmenter & Nizamudeen, 2021, Long *et al.*, 2022, Hayles & DAM, 2021, Li *et al.*, 2023) and the development of cryogenic shuttle transfer systems have significantly expanded the capacity of cryo-TEM (Wagner *et al.*, 2020, Huang *et al.*, 2022, Schaffer *et al.*, 2019).

Atom Probe Tomography (APT) has similarly benefited from the development of integrated hardware for cryogenic sample preparation (Stephenson *et al.*, 2018, Gerstl & Wepf, 2015, Stender *et al.*, 2022b, Perea *et al.*, 2017, Macauley *et al.*, 2021). APT provides three-dimensional compositional mapping with sub-nanometer resolution, and as such could provide complementary insights to cryo-TEM, yet its potential for studying biomacromolecules has remained almost untapped (Kelly *et al.*, 2012, Grandfield, 2022). Several reasons have prevented widespread adoption, including: (1) the lack of integrated cryo-preparation and transfer infrastructure for both cryo-APT and cryo-TEM, (2) different substrate types, geometries, and holder incompatibilities, and (3) difficulty extracting the cryo-TEM specimen grids and transferring them into APT (registration and marking).

Previously, biological APT samples have been successfully prepared from organics in non-liquid form (Prosa *et al.*, 2010, Proudian *et al.*, 2019) (Rusitzka *et al.*, 2018, Perea *et al.*, 2016). However, the development of APT samples from frozen liquids still remains in its infancy. In the past few years, methods including trapping thin liquid layers on a sharpened probe with graphene sheets (Qiu *et al.*, 2020a, Qiu *et al.*, 2020b), dropping water onto a substrate wire (Schwarz *et al.*, 2020) or nanoporous substrates (El-Zoka *et al.*, 2020, Perea *et al.*, 2020) have been developed. Other groups have also explored the possibility of using TEM grids for making ice specimens amenable to APT analysis (Zhang *et al.*, 2021a, Zhang *et al.*, 2021b, Zhang *et al.*, 2022).

Successful preparation of frozen liquid samples was achieved through the use of a nanoporous substrate, such as chemically dealloyed nanoporous Au (El-Zoka, 2018). Specimens were prepared by plasma FIB (El-Zoka et al., 2020), by making micropillars from a bulk surface with subsequent sharpening into APT specimens, following on the "moat" method (Miller *et al.*, 2005) for gallium FIB and "crater" method (Halpin *et al.*, 2019) for Xe plasma FIB (PFIB) . In our group's experience, this method has some practical limitations. Firstly, only areas on one edge of a bulk sample can be easily accessed for laser pulsing for APT specimen fabrication due to the positioning laser direction for measurement. Secondly, the volume of material removed practically requires a plasma FIB system, and it can lead to substantial redeposition, problems with subsequent annular milling steps (e.g. differential etching rates of redeposited material versus ice), and redeposited metal material residually redepositing on the final sharpened APT specimen. Thirdly, the large volume of liquid on a bulk substrate carried into the atom probe eventually leads to extensive contamination of the counter-electrodes, which requires their removal from the APT for cleaning. One alternative approach has been to make "matchsticks" of binary alloys, such as CuMn, and chemically dealloy to form nanoporous supports, which have water trapped in the pores (Tegg *et al.*, 2021).

In this paper, we explore the possibility of developing a "smart" substrate, that is both cost-effective and versatile in application. The material chosen was a Cu-Zn substrate, which is a low-cost material that provides consistent nanoporous dealloying results (Foroozan Ebrahimy, 2021, Ibrahim *et al.*, 2021). Versatility is achieved through designing a substrate that it is compatible with a TEM holder for purposes of achieving correlative microscopy between APT and TEM (Felfer *et al.*, 2012, Herbig *et al.*, 2015), and through minimizing the substrate bulk, eliminating issues of excess redeposition and counter-electrode contamination.

We introduce two new substrate geometries: (1) a grid with dimensions suitable for possible TEM/cryo-TEM examination; and (2) a self-aligning grid on the end of a support, which we term as the crown.  Jigs designed to simplify loading and handling of such grids and to facilitate the self-alignment of samples were also developed. Overall, the grid design was chosen to be compatible with a liquid ethane plunge freezing setup and to facilitate transfer between an atom probe, SEM-FIB, and TEM. These new designs provide a pathway to facilitate routine analyses of frozen liquids and biologically-relevant molecules in their native environment in the future.

## 2   Materials & Methods

### 2.1   Materials

Brass (Cu: 63%, Zn: 37% (atomic %), commonly known "yellow brass". XRD confirmed that as-received CuZn sheets were α-brass, see Supplementary Information (SI) **Error! Reference source not found.**.  Hereafter, we will refer to this material as "CuZn". It was selected for similarity to previous work, low cost, and commercial availability.  CuZn sheets of different thicknesses were purchased from Metall Ehrnsberger GbR (Metall Ehrnsberger GbR, Teublitz, Germany), and 100 µm thick foils for TEM grid fabrication purchased from LLT Applikations (LLT Applikations, Ilmenau, Germany).

Different annealing conditions were tried to remove the damage-affected zone, with optimized annealing conditions for CuZn found to be 600°C for 1 hour with furnace cooling, both under argon, with lower temperatures not fully removing the damage from the zone affected by the fabrication. All reported APT datasets were taken with CuZn annealed for 1 hour at 600 °C under argon with furnace cooling unless otherwise specified. Hydrochloric acid (HCl) (37%), phosphoric acid ($H_3PO_4$) (85%), sodium hydroxide (NaOH) (40%), and arginine hydrochloride powder were obtained from Sigma-Aldrich (Sigma-Aldrich, Munich, Germany). Type 1 ultrapure water was obtained either from Sigma-Aldrich or internally using an ultra-pure water system with resistivity greater than 18 MΩ-cm.  All solutions were prepared using Type I water, unless otherwise specified.

### 2.2   Equipment

TEM half grids (in disc form, for prototyping) were punched using a vertical TEM grid punch (CL 750 K, Mäder Pressen GmbH, Neuhausen, Germany).  Laser-cut TEM grids were manufactured via femtosecond laser cutting by LLT Applikations (LLT Applikations, Ilmenau, Germany). The crowns were prepared with electrical discharge machining (EDM) / spark erosion on a Mitsubishi EDM system.  The custom jigs for loading Felfer holders were fabricated by the MPIE mechanical workshop.

Where necessary, plasma cleaning was performed with an XEI Evactron C remote plasma cleaner at 10W forward power using air at a pressure of 0.5 Torr. Contact angle measurements were done with a Dataphysics Contact Angle System OCA and analysed with the Dataphysics software. Optical images of the grids were taken with a Zeiss Axioscope A1 optical microscope (Carl Zeiss Microscopy, Jena, Germany).

EBSD data was taken using an EDAX Hikari EBSD camera (EDAX Instruments, Pleasanton, CA, USA) on a Zeiss Sigma SEM (Carl Zeiss, Oberkochen, Germany) and processed using EDAX OIM 8.0 EBSD analysis software. XRD data was taken with a Bruker Diffractometer D8 Advance A25-X system and analysed with DIFFRAC EVA phase analysis software, version 4.3.0.1.  EDX data was taken using an EDAX ELECT PLUS EDX detector on a Helios PFIB and/or Helios 600i (detector area 30mm$^2$ area and 10mm$^2$ respectively) and processed with TEAM software.

The workflows for the APT experiments use the infrastructure described in Ref. (Stephenson et al., 2018).  A nitrogen-filled glovebox (Sylatech GmbH, Walzbachtal, Germany) was used for sample preparation (dew point -98°, oxygen level 20ppm, with actively switched catalyst beds for resorption) and loading into a vacuum transport suitcase. A FEI Xe-plasma FIB/SEM (Helios PFIB, FEI, Hillsboro, OR, USA), equipped with a custom intermediate chamber/cryogenic sample loading system (Microscopy Solutions LLC, Australia) and a Gatan C1001 cryogenic stage (Gatan Inc. Pleasanton, CA, USA), was used for most of the cryogenic sample milling and sharpening.  During the final part of experiments for this manuscript, the Gatan stage was removed and replaced with an Aquilos 2 cryo-stage, but the installation and sign-off was not completed before the completion of the manuscript.

Therefore, a Helios 5 CX gallium FIB (Thermo-Fisher) equipped with an Aquilos 2 cryogenic stage (Thermo-Fisher) was utilized for some for some final cryogenic specimen preparation and analysis. During operation, the Gatan cryogenic stage was set to an operating temperature of -180°C and the Aquilos 2 cryogenic stage was set to -190°C. The parameters used for the Gatan stage are provided in (El-Zoka et al., 2020), and for the Aquilos 2 cryo-stage in (Woods, 2023a). Additionally, two FEI gallium FIBs (Helios 600 and 600i) were used to make some cross-sectional SEM and ion images. Note that below, SEM figures are outlined in red and ion beam / FIB images will be outlined in blue. All other images (optical microscope images, CAD drawings, etc.) will be outlined in black.

Atom probe data was acquired using a straight flight path Cameca LEAP 5000XS atom probe (Cameca Instruments, Madison, WI, USA). Unless otherwise stated, all data was taken in laser mode using the following parameters: temperature 50K, 200kHz, 0.5% detection rate, laser power 40pJ, auto voltage mode (Rapid). Samples were transferred using a Ferrovac VCT100 Ultrahigh-Vacuum Cryogenic Transfer Module (UHVCTM) vacuum suitcase, hereafter "suitcase". All specimens analysed via APT were placed in Cameca APT cryogenic specimen holders with PEEK plastic thermal isolators, hereafter "cryo-pucks".

## 3   Method development

### 3.1   Nanoporous Copper TEM Grid

#### 3.1.1   Brass

The as-received material did not consistently chemically dealloy, which probably was a function of grain size, mechanical stress from fabrication and surface contamination. Here, EBSD shows the as-received CuZn had an average grain size of 5-7μm, see **Error! Reference source not found.**. The relationship between the grain size and the dealloying process was not investigated here in detail, but constitutes a possible future study.

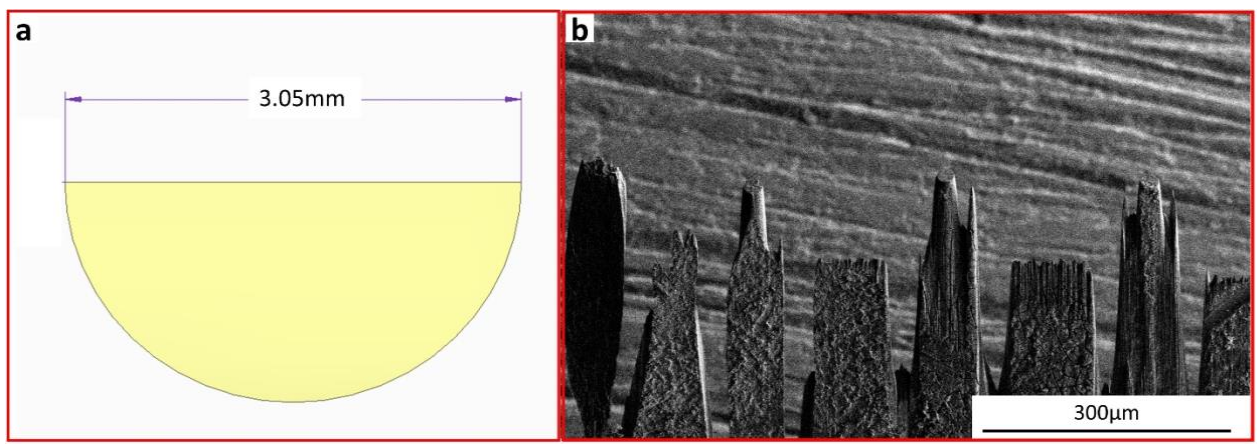

**Figure 1** (a) Prototype grid after TEM Punch (b) Grid after PFIB etching

### 3.2   Initial prototype

TEM-type CuZn half-discs (3.05mm diameter) for grid prototyping were punched out from 100μm thickness CuZn foil on the edge using a vertical TEM grid punch. The CuZn was initially etched for 2 hours in hydrochloric acid, as suggested in (Foroozan Ebrahimy, 2021). After etching, they were immersed in Type I ultra-pure DI water, rinsed twice, and left to soak for 1 hour, then placed in Felfer holders and left in Type I water overnight with the half-disc immersed. Since APT requires individual sharpened specimens with a minimum separation distance typically of the order of 400 μm, individual grid posts had to etched into the flat top of the punched half disc.  A sample half-disc is shown in Error! Reference source not found.**(a)**.  Prototype grids were created by etching individual posts with a flat top edge by using a plasma FIB which required approximately eight hours time (dependent on the number of posts etched), as in Error! Reference source not found.**(b)**.  While useful for prototyping, this was not a scalable solution to routinely produce these grids, and water was not well retained on the prototype, as shown in **Error! Reference source not found.**.

### 3.2.1   Design

A more consistent design was developed, which resemble traditional FIB half-grids for TEM. The final CAD grid design is shown below in Error! Reference source not found.**(a)** .  Half-grids with pre-cut spikes were produced by femtosecond laser cutting from a 100μm thickness foil in Error! Reference source not found.**(b)**.

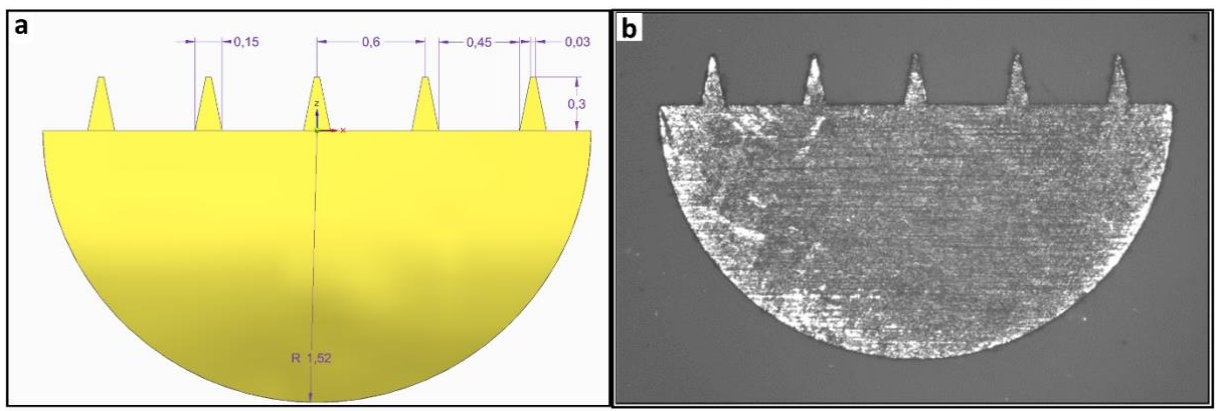

**Figure 2** (a) FIB half-grid CAD drawing, units in mm (b) Fabricated grid

As expected, the laser machining produced a heat-affected damage zone and high surface roughness, which was quite visible in the SEM after FIB cross-sectional cut, illustrated in Error! Reference source not found.**(a)**), and the grain structure is not well defined.  After annealing at 600˚C for 1h, the grain size significantly increased to 12.6μm, as shown in **Error! Reference source not found.**.  As shown in Error! Reference source not found.**(b)**, no damage zone is present below the surface roughness after FIB cross-section, and the grains below the surface are continuous. After annealing, there is still surface roughening, which has no effect on subsequent dealloying.

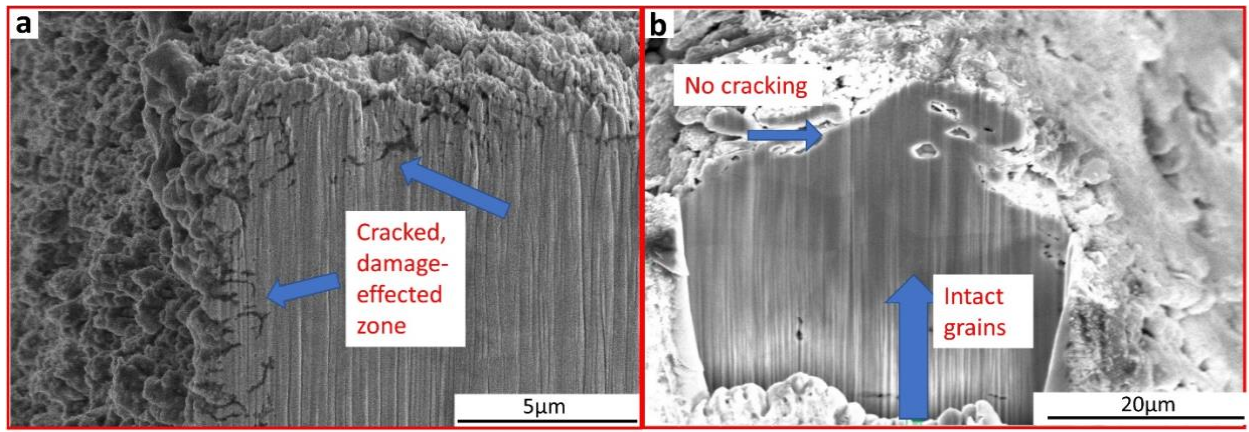

**Figure 3** (a) Cross-section, cracked damage-effected zone, as received after laser cutting (b) after heat treatment, no damage zone, large, visible grains

Previous work (Ibrahim et al., 2021) indicated that NaOH or HCl would both be potentially effective to dealloy CuZn alloys, but after experiments with a wide range of concentrations of NaOH and HCl (0.01M to 7M) for various dealloying times (1 – 96 hours), an optimal parameter set was found, in terms of surface roughening (via SEM), visual water retention after rinsing, etc. CuZn grids were dealloyed for 4 hours in 37% hydrochloric acid (HCl) and rinsed several times in Type I deionized water. Overall, the zinc concentration in a crown, where only the teeth are dipped into a stirred beaker of HCl for 4 hours, sees no composition change in cross-section (see composition in **Error! Reference source not found.** and surface image in **Error! Reference source not found.**), and a small change on the tip surface (see composition in **Error! Reference source not found.** and composition in **Error! Reference source not found.**). A SEM secondary electron image of the surface after CuZn dealloying is shown in Error! Reference source not found.**(a)**, illustrating increased porosity. A higher magnification image in Error! Reference source not found.**(b)** shows the nanoporous structure more clearly, illustrating that the presence of voids and that the contrast shown in **Error! Reference source not found.(a)** actually shows topographic changes versus shadowing or other SEM artifacts.

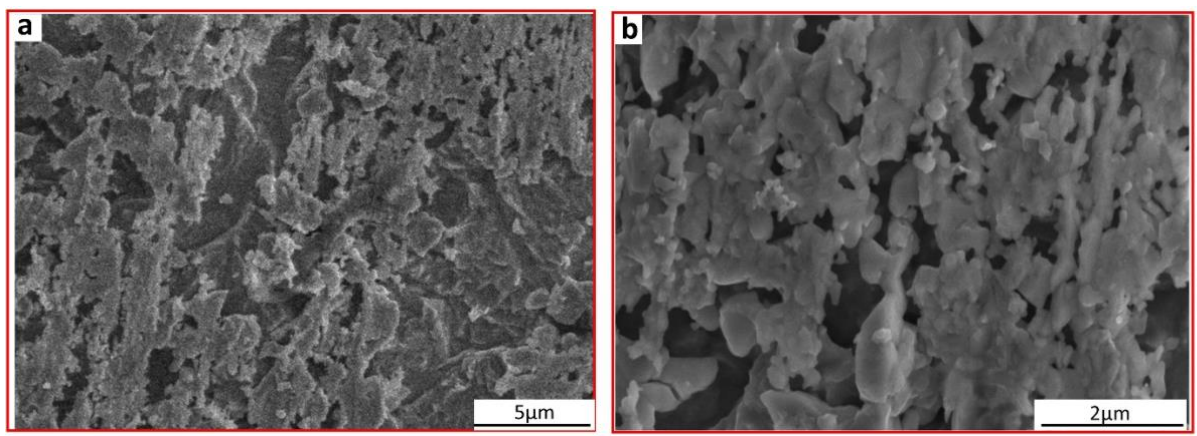

**Figure 4** (a) CuZn SEM secondary electron image after dealloying (b) Higher resolution secondary electron image

### 3.2.2 Grid mounting / sample preparation

Fabricated grids were mounted in Felfer holders using a custom-designed jig, shown in Error! Reference source not found.**(a)**, for ease of mounting.  This holder was inserted into a cryo-puck (Cameca) either in air, or under liquid nitrogen in the nitrogen-filled glovebox using the custom jig set shown in  Error! Reference source not found.**(b)**.  The CAD files for these holders are available publicly (Woods, 2023b).  The loaded cryo-puck assembly was typically placed upside down in a hole through a Parafilm-covered beaker (or suspended upside down in a custom holder with the teeth immersed into a bowl) containing the aqueous solution of interest to provide time for the material to get into the nanoporous structure, such that the end of the Felfer holder and the tips of the grid (and later crown) were immersed into the solution.  If excess liquid was present (very rarely), it was blotted such that a large drop would not block access to the tips. Once the grid had been wetted (or let soak), the Felfer holder (or crown) and cryo-puck were plunged into liquid nitrogen and transferred into the pre-cooled Ferrovac suitcase. The samples were then transferred to the PFIB under vacuum in the suitcase and mounted onto the pre-cooled Gatan cryo-stage.

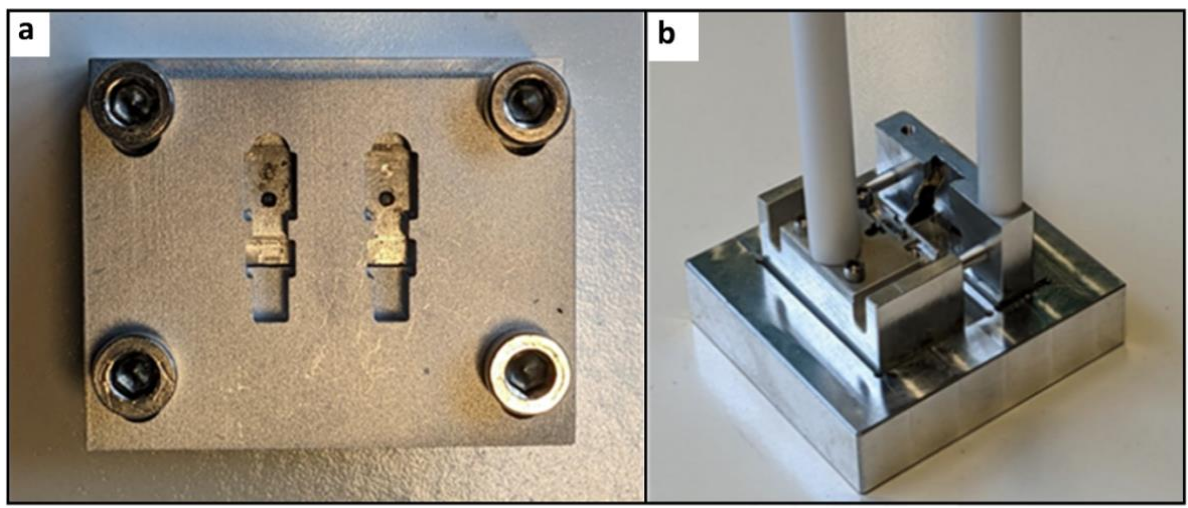

**Figure 5** (a) Felfer holder (TEM grid) loading jig (b) assembled multi-jig holder for aligning and transferring Felfer holders into atom probe specimen holders ("pucks")

Once loaded into the PFIB, the tips of the grids were examined to determine the amount of ice present. The most promising tips were selected and sharpened using progressively lower annular milling currents until they reached a final tip radius, e.g. less than or equal to 100nm, with ion beam currents as described in (El-Zoka et al., 2020). In addition, different strategies, such as cutting slots into the tip to increase water, such as discussed in (Zhang et al., 2022, Zhang, 2022), as well as arrays of holes, have been trialled for increasing the amount of retained water, with mixed success (see section Controlling Hydrophilicity). After sharpening, the sample on the cryo-puck was transported to the LEAP 5000 XS atom probe through the suitcase.

### 3.3 Brass Crown Design

#### 3.3.1 Design

Handling and mounting dealloyed half-grids into the holder without damaging them proved to be difficult, because they are brittle and prone to bending. Additional difficulties arose from misalignment in the Felfer holder, as mounting the grid in the glovebox was extremely challenging. Therefore, a unitary sample support made of CuZn was designed, hereafter called "crown," which was functionally equivalent to a half-grid held in a Felfer holder. The designs goals were to provide multiple tips (preferably at least 5, hereafter called "teeth"), which would hold liquid, and the crown needed to be self-supporting and self-aligning in the cryo-puck. No manual adjustments to the crown should be necessary, e.g. it could be mounted into the cryogenic puck first and no other actions would be necessary. The design initially had 1mm teeth length, along with a hole in the bottom to secure it in place with the attachment set screw in the front of the APT cryogenic puck holder.  Experiments showed that 1mm long tips were unevenly consumed by electropolishing (see Electropolishing section), so they were lengthened in the current version to 2mm.  The final CAD design is shown in Error! Reference source not found.**(a)** and an optical view of a polished crown in Error! Reference source not found.**(b)** respectively.  The as-received material was confirmed to be identical to previous batches, and fine polished using 0.05μm colloidal silica.  Crown thickness was an important factor - thinner CuZn metal crowns (0.4mm) faced issues with mechanical stability (including during APT data acquisition), and 0.7mm thick CuZn sometimes had mechanical vibrations with the helium compressor cooling the atom probe sample stage, whereas 0.9mm thick CuZn crowns did not.

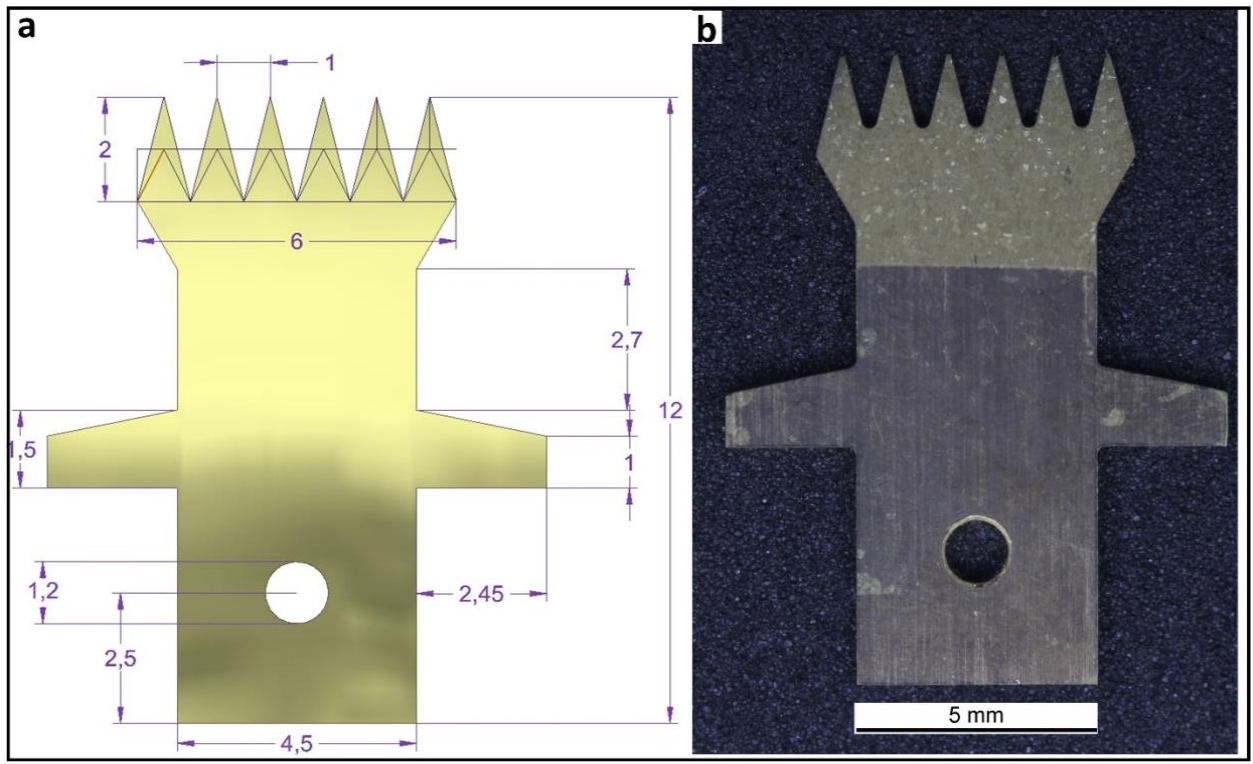

**Figure 6**. (a) Crown CAD design (b) Fabricated and polished crown with slight thickness variation from CAD design. Scalebar from optical microscope software. The slight difference in the size arises from fabrication tolerances, etc.

### 3.3.2   Crown Mechanical Polishing and Annealing
After EDM fabrication, polishing was performed using 200 and 500grit paper on a polishing wheel at 80rpm, which optimally yielded an end thickness of around 100μm.  However, much more reliable results can be achieved using tripod polishing, where three crowns can be attached via cyanoacrylate glue to the angled rod. After polishing with the same parameters, the metal rod was left overnight to soak. As a note, having too small a tip area could prove detrimental later, if the crowns are not handled carefully, and can be easily bent, even taking off the polishing wheel (see **Error! Reference source not found.**). Consistency greatly depended on operator proficiency and experience, and such thin tips are easily bent. The optical image shows the resultant polished area in side view (**Error! Reference source not found.(a)** shows an optical view, **Error! Reference source not found.(b)** shows an SEM overview). A holder box was modified to safely store the crowns upright and avoid contact that could lead to bending of the individual sharpened tips or "teeth". Polishing resulted in a mechanically damaged zone, so the crowns were annealed (600°C for 1 hour under argon with furnace cooling also under argon), which showed a similar change in grain size as observed after annealing to remove the damage zone. Subsequent cross-sections demonstrated elimination of the mechanically damaged zone, as previously described for the CuZn half-grids.

## 4 Results & discussion

### 4.1 APT specimen preparation

Once crowns have been fabricated, dealloyed, dipped into water or solutions, plunged into liquid nitrogen, and inserted into the PFIB, the individual water-bearing tips or "teeth" can be sharpened into APT-suitable specimens using a similar approach to what is described in Ref. (El-Zoka et al., 2020). A secondary electron image of a PFIB-etched vacuum-dealloyed crown specimen is displayed in **Error! Reference source not found.(a)**, with electron charging artifacts at the top of the image due to the presence of water. In **Error! Reference source not found.(b)**, a thin layer of water on top of a "tooth" is readily visible, with the water appearing darker as it emits less backscatter electrons.

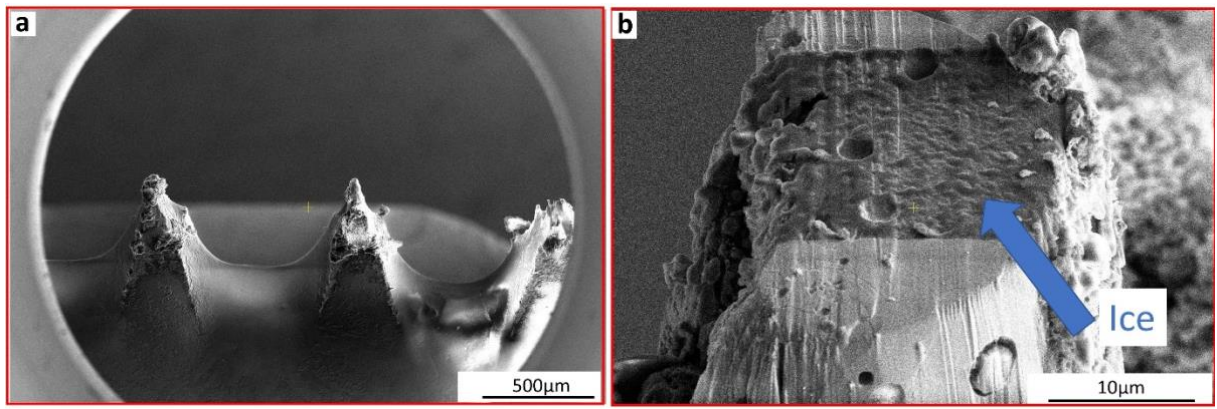

**Figure 7** (a) Secondary electron image of the top of a crown obtained at cryogenic temperature (b) partially etched "tooth" with thin residual ice

## 4.2 APT analyses

**Error! Reference source not found.(a)** plots the mass spectrum from the analysis of a 0.1M arginine hydrochloride solution deposited on brass without dealloying (taken at 60pJ laser power), with the specimen shown in **Error! Reference source not found.(b)**, and the corresponding 3D reconstruction displayed in Error! Reference source not found.**(c)**. Initially, only short datasets could be obtained, and the layer of retained solution was typically very thin, as illustrated in **Error! Reference source not found.(b)**, highlighting the importance of controlling the hydrophilicity of the surface as suggested in (Stender *et al.*, 2022a). In comparison to the nanoporous gold reported previously, a multitude of peaks appear created by the use of the CuZn alloy, with numerous peaks corresponding to Cu and Zn oxides and hydroxides, along with mixed molecular ions containing both cations (e.g. $ZnO_x$, $CuZnO_x$, etc.). A close-up on the mass spectrum in the range of 85-145 Da, illustrates how these mass peaks at every Da interval, as shown in **Error! Reference source not found.**, would strongly interfere with the detection of organic fragments.

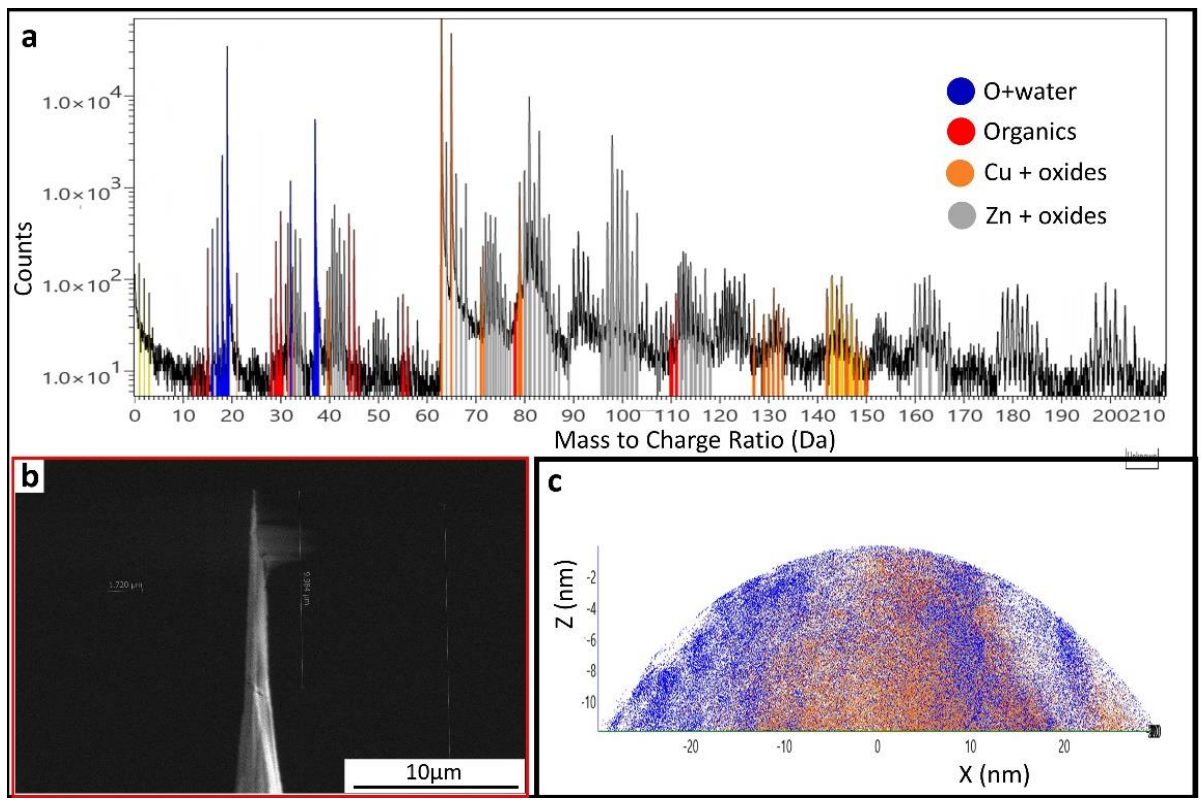

**Figure 8** (a) Mass Spectrum for Water on CuZn support (b) SEM image of Ice tip on brass (c) 3D APT tip view

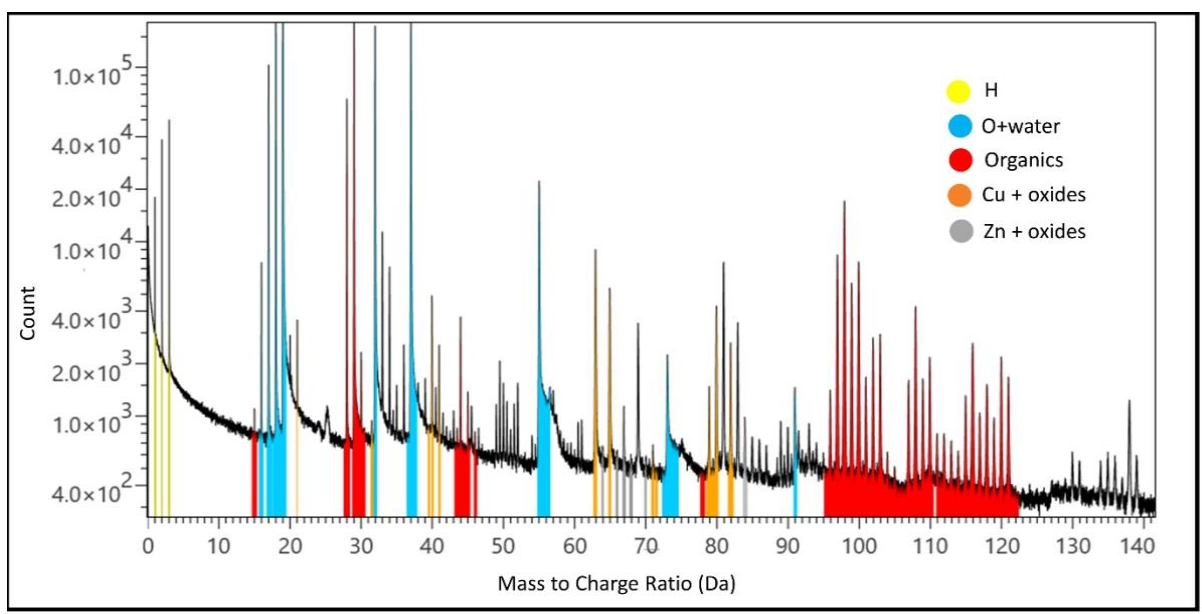

**Figure 9** Mass spectrum of amino acid on dealloyed brass substrate, with less than 2% copper by atomic percent, with minimal zinc

APT analyses obtained from chemically dealloyed substrates produced longer datasets with relatively fewer but still numerous spurious peaks pertaining to Zn and Cu oxides (see **Error!**

**Reference source not found.**, data taken at 70pJ). The brass crown had been immersed for 4h in HCl for dealloying. As an alternative, we have explored vacuum dealloying for CuZn, e.g. annealing brass pieces in a holder open to high vacuum (e.g. $10^{-5}$ mBar) at 800°C for one hour, with vacuum cooling. We finally opted for this approach, as it produces surfaces with much less Zn, below 1 at% from EDS measurement, in comparison to chemical dealloying (see **Error! Reference source not found.** for composition and **Error! Reference source not found.** for Cu and Zn spatial distribution maps). When residual Zn remains near the sample's surface and the chemically or vacuum dealloyed CuZn is left in water for some time, it can become potentially much more concentrated and thusly problematic, which will be discussed in more details below. An example of a high quality mass spectrum of an frozen aqueous solution, namely 0.1M arginine HCl acquired at 70pJ, is shown in **Error! Reference source not found.**, and the composition by as calculated by APT bulk composition in AP Suite contains less than 2 at% Cu and only traces of Zn (see **Error! Reference source not found.**).

## 4.3 Dealloying

### 4.3.1 Chemical Dealloying

It is critical to ensure that as much of the Zn as possible is removed from the surface region of the teeth of the grid or crown. It appears that if Zn is present in more than 1–2% atomic percent, a wide array of surface hydroxides and oxides are detected. These create series of peaks at 1Da interval over a large range of mass-to-charge ratios. These mass interferences render the mass spectrum and APT measurements effectively useless, as Figure 7 illustrates, because peaks cannot be reliably assigned to molecular ions from organic fragments. Thorough Zn removal can be difficult to achieve using chemical dealloying means. Even with optimized dealloying conditions using HCl, pockets of Zn can be present in the substrate material, which can effect the solution to be analyzed. With respect to the dealloying times, if thin CuZn TEM grids are left in HCl for too long (although roughness does increases as shown in **Error! Reference source not found.**(a)-(d) for a 3x7x0.1mm CuZn piece), they crack and become too brittle to mechanically handle (as shown in **Error! Reference source not found.**), or simply uncontrollably dissolve.

As **_Error! Reference source not found._**, shows, the $O_2^+$ peak at mass 32 Da, possible residual $Zn^{2+}$ at masses 33 Da and 34 Da, miscellaneous organic masses, and $(H_2O)_2H^+$ cluster with mass 37 Da along with inherent thermal tails would mask any residual $Cl^+$ or make it exceptionally difficult to definitively identify any mass peak at 35 or 37 Da. The chemical dealloying effects do enhance the grain structure optically, as in **Error! Reference source not found.**, but the overall roughening induced is not visible. It is critical to note that larger samples (e.g. 3 x 7 x 0.5mm for example) left in 37% HCl for 4 hours (no matter the volume) will only be slightly Zn-depleted at the surface (e.g. composition will shift from 63% Cu, 37% Zn to 70% Cu, 30% Zn when measured via EDX). This ratio is consistent whether measured over square or rectangular area sizes ranging from 20μm to over 1mm per side.

### 4.3.2  Vacuum Dealloying

Vacuum dealloying consistently yielded roughened copper surfaces with less than 1% atomic percent Zn as per EDX mapping, as illustrated by composition measurements in **Error! Reference source not found.** and spatial distribution in **Error! Reference source not found.**. These measurements were taken across several samples and quantified as 1-1.5% atomic percent via APT.  Although there were small pockets with somewhat higher composition (5-6% atomic percent via APT), based on series of atom probe analyses from close-by regions, a rough estimate of their size is approximately 1–3μm. These pockets are also rare – a rough estimate from our EDS mapping was below 1% by area. The vacuum dealloying technique drastically increased the grain size to an average of 80μm (including edge grains, **Error! Reference source not found.**) or larger (including edge grains, **Error! Reference source not found.**).  The surface, after removal from the annealing furnace, exhibited increased roughness in SEM images and natural hydrophilicity. This method provides a promising alternative to chemical means and leaves very little Zn in APT analysis, at least to the extent that any peaks associated to hydroxides and oxides are not discernible above background in APT.  Residual Zn seem to be concentrated in pores that are between 10-20μm, which appear to concentrate the Zn.

### 4.4  Controlling Hydrophilicity

One of the most significant problems is that the water tends to form a large droplet around the bases of the "teeth" on both the crowns and the grids and the sharpened tips retain only a small amount of water at the top.  Therefore, finding the right tooth diameter is important, as is ensuring that their top remains relatively flat to hold a sufficient volume of water. However, FIB cutting the tips flat– after dealloying – makes them bad at retaining water, with results as shown in Error! Reference source not found.**(b).**  The flat cutting must be done prior to dealloying, which even does not yield consistent results.

Surface energy, as estimated from the contact angle, is an important factor in wetting. When a flat brass or copper sample is stored in a plastic container, after some time it adsorbs organics on its surface, which creates a hydrophobic surface with an as- measured 110° contact angle. Five minutes of plasma cleaning using a remote plasma (e.g. air, conditions specified in Methods) restores the measured contact angle to 50°, restoring hydrophilicity.

Some grid designs with arrays of slots or holes were initially prototyped, following previous reports by (Zhang, 2022) . An example with slots is shown in **Error! Reference source not found.(a)**. Though water did localize in the slots, as shown in **Error! Reference source not found.(b)**, these slots did not produce any significant benefit for water retention, with similar results for holes in **Error! Reference source not found.**(c) and lack of water in **Error! Reference source not found.**(d). Ultimately, even if one could successfully use this approach to control the hydrophilicity of the surface, a key issue will arise when shaping APT specimens, because of the differential ion milling rate of brass versus water. When tried with the chemically dealloyed grids, this made retention of the water difficult, with visible differences between Ga and Xe FIBs, making controlling this process extremely challenging for routine preparation. Although with this work mostly used large diameter teeth as prototypes, most larger teeth diameters (e.g. 30–90μm) are only practical with the use of a Xe plasma FIB, which can etch large amounts of material.

## 4.5   Electropolishing

Electropolishing with phosphoric acid was also attempted in order to reduce the initial size of the teeth, so as to reduce the beam time necessary on the FIB. However, the non-symmetrical designs like the crowns – and the grids – made the control of the electropolishing process more challenging, and led to non-optimal and inconsistent results. Tips farther from the edges etched more than those in the middle, even after the crown design was changed to have 2mm length "teeth" from original 1mm length. Please see Supplemental Information 8.1.1 "**Error! Reference source not found.** for methods, discussions, and results.

## 4.6   Future Work

Smaller tip diameters, which are compatible with usage in gallium FIBs, need to be optimized. First, there is an observed practical limit to tip diameter – when the diameter becomes too small, water does not stay on the tip, and this will need to be quantified in the future. Secondly, from a practical perspective, manufacturing tips with flat tops but small diameters are technically demanding. The short-term feasible solution seems to be to use a 90° holder and using the FIB to tilt the ends of the crowns perpendicular to the ion beam. After that, annealing the crowns at 800°C under vacuum as described previously to dealloy them should remove mechanical damage and make them hydrophilic, which currently produces inconsistent results (e.g. poor wetting like Error! Reference source not found.**(b)).** The final procedure, including cutting or polishing the tips to make them flat, or subsequently roughening them, must be optimized to produce consistent results. Another option is to use laser micro-machining to create conical tips of varying diameters, which is being investigated.

New crown versions with different size flat-topped tips are envisioned to be tested, which should retain water more strongly, and be easier to handle, probably fabricated with laser cutting. Additionally, changing the "teeth" spacing to 0.7mm wide, e.g. making the width of the wedge symmetric with the crown thickness, may improve the electropolishing results. Improved electropolishing techniques are under development, which will involve further optimizing acid concentration and voltage, as well as optimizing dipping time into the solution.

The effect of surface treatments (argon and oxygen plasma cleaning) to control surface hydrophilicity will be further investigated, since even thin surface oxides can strongly influence the contact angle, as can adsorbed amorphous carbonaceous material if stored in plastic containers. Sample storage in glass containers will be used. Preliminary results show that this is a promising way to ensure consistency and repeatability in tip wetting, providing it is done quickly before the use of the tips.

## 5   Conclusion

Two designs of carrier substrates for the preparation of aqueous solutions for APT using plasma FIB systems have been presented here, along with workflows showing how they can be used to produce atom probe tips. Both avoid the need to perform cryogenic lift-out or the use of a cryogenic micromanipulator in the FIB system. The use of a cheap, widely commercially available substrate material has been demonstrated, and a design which can be easily fabricated in dozens or hundreds of pieces via EDM has been shown, as well as successful mechanical polishing of such samples, several at a time. These results show that successful sample preparation is possible for both configurations.

## 6 Conflicts of Interest

There are no conflicts to declare.

## 7 Acknowledgments

EVW, SHK, AEZ, and BG are grateful for funding from the ERC for the project SHINE (ERC-CoG) #771602. BG is grateful to the DFG for funding through the Leibniz Award. We thank Uwe Tezins, Christian Broß and Andreas Sturm for their support at the FIB and APT facilities at MPIE. We would like to thank Katja Angehangt, Monika Nellessen, and Christian Broß for their assistance with sample preparation. We would like to thank Tristan Wickfeld, Rainer Lück, Ralf Selbach, Mario Bütow, and the entire MPIE mechanical workshop for their assistance with design and fabrication of the jigs and crowns. We would like to thank Jürgen Wichert for his kind assistance with sample annealing.

## References

Cheng, Y., Grigorieff, N., Penczek, P. A. & Walz, T. (2015) A primer to single-particle cryo-electron microscopy. *Cell*, **161,** 438-449.

Doerr, A. (2016) Single-particle cryo-electron microscopy. *Nat Methods*, **13,** 23.

Dubochet, J., Adrian, M., Chang, J. J., Homo, J. C., Lepault, J., McDowall, A. W. & Schultz, P. (1988) Cryo-electron microscopy of vitrified specimens. *Q Rev Biophys*, **21,** 129-228.

El-Zoka, A. (2018) Characterization and Functional Improvement of Nanoporous Metals. In: *Chemical Engineering and Applied Chemistry.* University of Toronto, Toronto.

El-Zoka, A. A., Kim, S. H., Deville, S., Newman, R. C., Stephenson, L. T. & Gault, B. (2020) Enabling near-atomic-scale analysis of frozen water. *Sci Adv*, **6**.

Felfer, P. J., Alam, T., Ringer, S. P. & Cairney, J. M. (2012) A reproducible method for damage-free site-specific preparation of atom probe tips from interfaces. *Microsc Res Tech*, **75,** 484-491.

Foroozan Ebrahimy, A. (2021) Metallic Nanoporous Materials, from Design to Degradation In: *Chemical Engineering and Applied Chemistry.* University of Toronto Toronto.

Gerstl, S. S. A. & Wepf, R. (2015) Methods in Creating, Transferring, & Measuring Cryogenic Samples for APT. *Microscopy and Microanalysis*, **21,** 517-518.

Grandfield, K., Micheletti, C., Deering, J., Arcuri, G., Tang, T., Langelier, B. (2022) Atom probe tomography for biomaterials and biomineralization. *Acta Biomaterialia*, **148,** 44-60.

Halpin, J. E., Webster, R. W. H., Gardner, H., Moody, M. P., Bagot, P. A. J. & MacLaren, D. A. (2019) An in-situ approach for preparing atom probe tomography specimens by xenon plasma-focussed ion beam. *Ultramicroscopy*, **202,** 121-127.

Hayles, M. F. & DAM, D. E. W. (2021) An introduction to cryo-FIB-SEM cross-sectioning of frozen, hydrated Life Science samples. *J Microsc*, **281,** 138-156.

Herbig, M., Choi, P. & Raabe, D. (2015) Combining structural and chemical information at the nanometer scale by correlative transmission electron microscopy and atom probe tomography. *Ultramicroscopy,* **153,** 32-39.

Huang, G., Zhan, X., Zeng, C., Zhu, X., Liang, K., Zhao, Y., Wang, P., Wang, Q., Zhou, Q., Tao, Q., Liu, M., Lei, J., Yan, C. & Shi, Y. (2022) Cryo-EM structure of the nuclear ring from Xenopus laevis nuclear pore complex. *Cell Res,* **32,** 349-358.

Ibrahim, S., Dworzak, A., Crespo, D., Renner, F. U., Dosche, C. & Oezaslan, M. (2021) Nanoporous Copper Ribbons Prepared by Chemical Dealloying of a Melt-Spun ZnCu Alloy. *The Journal of Physical Chemistry C,* **126,** 212-226.

Kelly, T. F., Nishikawa, O., Panitz, J. A. & Prosa, T. J. (2012) Prospects for Nanobiology with Atom-Probe Tomography. *MRS Bulletin,* **34,** 744-750.

Klumpe, S., Kuba, J., Schioetz, O. H., Erdmann, P. S., Rigort, A. & Plitzko, J. M. (2022) Recent Advances in Gas Injection System-Free Cryo-FIB Lift-Out Transfer for Cryo-Electron Tomography of Multicellular Organisms and Tissues. *Microscopy Today,* **30,** 42-47.

Li, S., Wang, Z., Jia, X., Niu, T., Zhang, J., Yin, G., Zhang, X., Zhu, Y., Ji, G. & Sun, F. (2023) ELI trifocal microscope: a precise system to prepare target cryo-lamellae for in situ cryo-ET study. *Nat Methods,* **20,** 276-283.

Long, D. M., Singh, M. K., Small, K. A. & Watt, J. (2022) Cryo-FIB for TEM investigation of soft matter and beam sensitive energy materials. *Nanotechnology,* **33**.

Macauley, C., Heller, M., Rausch, A., Kummel, F. & Felfer, P. (2021) A versatile cryo-transfer system, connecting cryogenic focused ion beam sample preparation to atom probe microscopy. *PLoS One,* **16,** e0245555.

Mahamid, J., Pfeffer, S., Schaffer, M., Villa, E., Danev, R., Cuellar, L. K., Forster, F., Hyman, A. A., Plitzko, J. M. & Baumeister, W. (2016) Visualizing the molecular sociology at the HeLa cell nuclear periphery. *Science,* **351,** 969-972.

Miller, M. K., Russell, K. F. & Thompson, G. B. (2005) Strategies for fabricating atom probe specimens with a dual beam FIB. *Ultramicroscopy,* **102,** 287-298.

Nannini, J. B. (2017) The Nobel Prize in Chemistry 2017. Nobel Prize Outreach AB.

Parmenter, C. D. & Nizamudeen, Z. A. (2021) Cryo-FIB-lift-out: practically impossible to practical reality. *J Microsc,* **281,** 157-174.

Perea, D. E., Gerstl, S. S. A., Chin, J., Hirschi, B. & Evans, J. E. (2017) An environmental transfer hub for multimodal atom probe tomography. *Adv Struct Chem Imaging,* **3,** 12.

Perea, D. E., Liu, J., Bartrand, J., Dicken, Q., Thevuthasan, S. T., Browning, N. D. & Evans, J. E. (2016) Atom Probe Tomographic Mapping Directly Reveals the Atomic Distribution of Phosphorus in Resin Embedded Ferritin. *Sci Rep,* **6,** 22321.

Perea, D. E., Schreiber, D. K., Ryan, J. V., Wirth, M. G., Deng, L., Lu, X., Du, J. & Vienna, J. D. (2020) Tomographic mapping of the nanoscale water-filled pore structure in corroded borosilicate glass. *npj Materials Degradation,* **4**.

Prosa, T. J., Keeney, S. K. & Kelly, T. F. (2010) Atom probe tomography analysis of poly(3-alkylthiophene)s. *J Microsc,* **237,** 155-167.

Proudian, A. P., Jaskot, M. B., Diercks, D. R., Gorman, B. P. & Zimmerman, J. D. (2019) Atom Probe Tomography of Molecular Organic Materials: Sub-Dalton Nanometer-Scale Quantification. *Chemistry of Materials,* **31,** 2241-2247.

Qiu, S., Garg, V., Zhang, S., Chen, Y., Li, J., Taylor, A., Marceau, R. K. W. & Fu, J. (2020a) Graphene encapsulation enabled high-throughput atom probe tomography of liquid specimens. *Ultramicroscopy,* **216,** 113036.

Qiu, S., Zheng, C. X., Garg, V., Chen, Y., Gervinskas, G., Li, J., Dunstone, M. A., Marceau, R. K. W. & Fu, J. (2020b) Three-Dimensional Chemical Mapping of a Single Protein in the Hydrated State with Atom Probe Tomography. *Anal Chem,* **92,** 5168-5177.

Rusitzka, K. A. K., Stephenson, L. T., Szczepaniak, A., Gremer, L., Raabe, D., Willbold, D. & Gault, B. (2018) A near atomic-scale view at the composition of amyloid-beta fibrils by atom probe tomography. *Sci Rep,* **8,** 17615.

Schaffer, M., Mahamid, J., Engel, B. D., Laugks, T., Baumeister, W. & Plitzko, J. M. (2017) Optimized cryo-focused ion beam sample preparation aimed at in situ structural studies of membrane proteins. *J Struct Biol,* **197,** 73-82.

Schaffer, M., Pfeffer, S., Mahamid, J., Kleindiek, S., Laugks, T., Albert, S., Engel, B. D., Rummel, A., Smith, A. J., Baumeister, W. & Plitzko, J. M. (2019) A cryo-FIB lift-out technique enables molecular-resolution cryo-ET within native Caenorhabditis elegans tissue. *Nat Methods,* **16,** 757-762.

Schwarz, T. M., Weikum, E. M., Meng, K., Hadjixenophontos, E., Dietrich, C. A., Kastner, J., Stender, P. & Schmitz, G. (2020) Field evaporation and atom probe tomography of pure water tips. *Sci Rep,* **10,** 20271.

Stender, P., Gault, B., Schwarz, T. M., Woods, E. V., Kim, S. H., Ott, J., Stephenson, L. T., Schmitz, G., Freysoldt, C., Kastner, J. & El-Zoka, A. A. (2022a) Status and Direction of Atom Probe Analysis of Frozen Liquids. *Microsc Microanal*, 1-18.

Stender, P., Solodenko, H., Weigel, A., Balla, I., Schwarz, T. M., Ott, J., Roussell, M., Joshi, Y., Duran, R., Al-Shakran, M., Jacob, T. & Schmitz, G. (2022b) A Modular Atom Probe Concept: Design, Operational Aspects, and Performance of an Integrated APT-FIB/SEM Solution. *Microsc Microanal*, 1-13.

Stephenson, L. T., Szczepaniak, A., Mouton, I., Rusitzka, K. A. K., Breen, A. J., Tezins, U., Sturm, A., Vogel, D., Chang, Y., Kontis, P., Rosenthal, A., Shepard, J. D., Maier, U., Kelly, T. F., Raabe, D. & Gault, B. (2018) The Laplace Project: An integrated suite for preparing and transferring atom probe samples under cryogenic and UHV conditions. *PLoS One,* **13,** e0209211.

Tegg, L., McCarroll, I., Sato, T., Griffith, M. & Cairney, J. (2021) Nanoporous metal tips as frameworks for analysing frozen liquids with atom probe tomography. *Microscopy and Microanalysis,* **27,** 1512-1513.

Tonggu, L. & Wang, L. (2020) Cryo-EM sample preparation method for extremely low concentration liposomes. *Ultramicroscopy,* **208,** 112849.

Wagner, F. R., Watanabe, R., Schampers, R., Singh, D., Persoon, H., Schaffer, M., Fruhstorfer, P., Plitzko, J. & Villa, E. (2020) Preparing samples from whole cells using focused-ion-beam milling for cryo-electron tomography. *Nat Protoc,* **15,** 2041-2070.

Walther, P., Chen, Y., Pech, L. L. & Pawley, J. B. (1992) High-resolution scanning electron microscopy of frozen-hydrated cells. *J Microsc,* **168,** 169-180.

Woods, E. V., Singh, Mahander P., Kim, Se-Ho, Douglas, James O., El-Zoka, Ayman, Guiliani, Finn, Gault, B. (2023a) A versatile and reproducible cryo-sample preparation method for atom probe studies. **arXiv:2303.18048**.

Woods, E. V. S. (2023b) Felfer holder CAD parts.

Zhang, S. (2022) Fabrication and Characterization of Nanoscale Needle Specimens in a Cryogenic Environment. Monash University

Zhang, S., Garg, V., Gervinskas, G., Marceau, R. K. W., Chen, E., Mote, R. G. & Fu, J. (2021a) Ion-Induced Bending with Applications for High-Resolution Electron Imaging of Nanometer-Sized Samples. *ACS Applied Nano Materials,* **4,** 12745-12754.

Zhang, S., Gervinskas, G., Liu, Y., Marceau, R. K. W. & Fu, J. (2021b) Nanoscale coating on tip geometry by cryogenic focused ion beam deposition. *Applied Surface Science,* **564**.

Zhang, S., Gervinskas, G., Qiu, S., Venugopal, H., Marceau, R. K. W., de Marco, A., Li, J. & Fu, J. (2022) Methods of Preparing Nanoscale Vitreous Ice Needles for High-Resolution Cryogenic Characterization. *Nano Lett,* **22,** 6501-6508.